# Critical Fields and Critical Currents in MgB$_2$

A. D. Caplin, Y. Bugoslavsky, L.F. Cohen, L. Cowey, J. Driscoll, J. Moore, and

G.K. Perkins

Centre for High Temperature Superconductivity, Blackett Laboratory,

Imperial College, London SW7 2BZ, UK

ABSTRACT

We review recent measurements of upper ($H_{c2}$) and lower ($H_{c1}$) critical fields in clean single crystals of MgB$_2$, and their anisotropies between the two principal crystallographic directions. Such crystals are far into the "clean limit" of Type II superconductivity, and indeed for fields applied in the $c$-direction, the Ginzburg-Landau parameter $\kappa$ is only about 3, just large enough for Type II behaviour.

Because $\mu_0 H_{c2}$ is so low, about 3 T for fields in the $c$-direction, MgB$_2$ has to be modified for it to become useful for high-current applications. It should be possible to increase $H_{c2}$ by the introduction of strong electron scattering (but because of the electronic structure and the double gap that results, the scatterers will have to be chosen carefully). In addition, pinning defects on a scale of a few nm will have to be engineered in order to enhance the critical current density at high fields.

I.  INTRODUCTION

The discovery of superconductivity in $MgB_2$ in early 2001 appeared to offer a material with an attractively high transition temperature $T_c$, but one without the major disadvantage of the high temperature superconductors (HTS): the opacity of HTS grain boundaries to supercurrent. To be useful for high current conductor applications, the critical current density $J_c$ needs to be above $\sim 10^4$ A cm$^{-2}$, and in most contexts that value has to be sustained in magnetic fields $B$ up to several Tesla. Furthermore, for $MgB_2$ to be competitive against conventional low temperature superconductors (LTS), the operating temperature $T$ has to be within reach of relatively simple (and inexpensive) cryocoolers, and so no lower than about 20 K.

Early measurements of $J_C(B,T)$ on standard commercial material demonstrated that indeed $J_C$ is very high in $MgB_2$, but only at relatively low temperatures and low fields. Disappointingly, $J_C$ drops off rapidly with increasing field (figure 1).[1] In the HTS context, suspicion would have fallen on grain boundary limitations, but detailed analysis of the $MgB_2$ magnetisation data suggested that even at high fields, current continued to circulate through the entire sample, and that the limiting factor was lack of vortex pinning. It was noted too that the irreversibility field $H_{irr}$ (defined as the field at which $J_C$ drops below some small, but arbitrary criterion, e.g. $10^3$ A cm$^{-2}$) in bulk material was only about one-half of the upper critical field $H_{c2}$, as inferred from resistive transitions. On the other hand, thin films showed much higher values of $H_{irr}$, both in absolute terms, and as a fraction of $H_{c2}$.[2] The latter was apparently also enhanced over that of the usual bulk material. Films grown by some techniques, but not all, suffer a significant reduction of $T_c$.[3]

The impact of disorder on $J_C(B,T)$ has been studied directly by looking at proton-irradiated samples.[4] The induced damage is in the form of point defects, and at a level of $\sim 1\%$ displacements-per-atom (dpa), the rate of depression of $J_C$ by applied field is substantially reduced. As with some of the thin films, there is a reduction in $T_c$, which perhaps accounts for the fact that at low fields, the damage depresses $J_C$.

All of this early work was conducted on polycrystalline samples, so that although some estimates were made of critical fields and anisotropy, the uncertainties were considerable. However, the normal state bulk resistivity of nominally-pure $MgB_2$ tends to be rather low, around 1 µΩ cm or even less just above the superconducting transition,[5] and this translates (using free electron parameters) to an electron mean free path Λ of order 100 nm. The coherence length ξ may be related simply to $H_{c2}$ from the number density and area of the vortex cores: $\mu_0 H_{c2} = \Phi_0 / 2\pi\xi^2$, where $\Phi_0$ is the flux quantum; if ξ is measured in nm, $\mu_0 H_{c2} \approx 300/\xi^2$ Tesla. In terms of the early estimated values of $\mu_0 H_{c2}(0)$ of order 10 Tesla, corresponding to ξ~5 nm, it was clear that in such samples, Λ>>ξ, and consequently as-prepared bulk polycrystalline "pure" $MgB_2$ is generally in the "clean limit" of Type II superconductivity.

As is well-known, strong elastic scattering does not of itself affect the thermodynamic parameters ($T_c$, the gap Δ, or the thermodynamic critical field $H_c$), but, because the electrons then diffuse rather than move ballistically, it does shorten the coherence length. In the extreme dirty limit, $\xi_{dirty} \rightarrow \sqrt{(\xi_{clean} \Lambda_{dirty})}$, and so strong scattering raises $H_{c2}$ by a factor of order $\xi_{clean}/\Lambda_{dirty}$.

Obviously, an important goal has been to raise $J_C(B,T)$ in $MgB_2$, and several groups have succeeded in doing so in bulk material by a variety of approaches. There are two distinct potential gains to be made, enhancement of $H_{c2}$, and strengthened pinning, but it has been unclear how much is attributable to each. Or, to put it the other way around, we have had little idea of how much headroom is available for increase of $H_{c2}$, and how much for stronger pinning.

## II. CRITICAL FIELDS OF MgB$_2$

It is only very recently that crystals of sufficient size and quality have become available to allow measurement of the critical fields, both lower and upper, of MgB$_2$. Even so, these crystals are small (and perhaps as with HTS crystals there may be an inverse relationship between size and quality), and the experiments are non-trivial.

### II.A. Measurement of H$_{c2}$

To establish $H_{c2}$ reliably, it is essential to ensure that what is being observed is the (reversible) extinction of superconductivity throughout the volume of the sample. The heat capacity jump at the transition provides the greatest confidence, followed closely by magnetic measurements. Resistive monitoring of the transition may be misleading, as the sample can be inhomogeneous, or indeed surface superconductivity may survive to fields up to $\sim\sqrt{3}H_{c2}$. Also, if the pinning is very weak, flux may move so easily as the field nears $H_{c2}$ that the resistivity approaches closely that of the normal state, and the transition becomes too rounded for accurate determination.

The heat capacity of MgB$_2$ crystals has recently been measured for field orientations both parallel and perpendicular to the *c*-axis, using a specialised rig of sufficient sensitivity to cope with microgram samples.[6] Magnetic techniques (SQUID, torque and vibrating sample magnetometers) are more widely available, and in principle have noise levels low enough to monitor the magnetic transition. Close to $H_{c2}$, the sample magnetisation *M* of a simple Type II superconductor is approximately:

$$-M = \frac{(H_{c2} - H)}{2\kappa^2} \tag{1}$$

where κ is the usual Ginzburg-Landau (G-L) parameter; this form is valid in the limit of very large κ. However, the abrupt change of slope of *M(H)* at $H_{c2}$ has to be located against a field- and temperature-dependent background from the sample holder etc., and so is not straightforward (Figure 2(a)).[7] As the name implies, torque magnetometers respond (with great sensitivity) to the sample magnetic moment with a signal proportional to $\mathbf{m}\wedge\mathbf{H}$, which necessarily vanishes along symmetry directions, so that the angular dependence of $H_{c2}$ has to be extrapolated to the principal directions. Furthermore, at temperatures approaching $T_c$, where $H_{c2}$ is small, torque magnetometers become less useful.

To overcome these difficulties we have adapted our scanning Hall imaging device[8] to look directly at the stray induction generated by the crystal magnetisation. The Hall probe alternates, at a frequency of ~0.1 Hz, between a position close to the sample, and one a few sample dimensions away from it (where the stray induction is negligible); the synchronous signal therefore arises solely from the sample magnetic moment. This technique yields an extremely clean trace of the sample magnetisation (figure 2(b)) that is seen to be reversible.[7] The amplitude of the signal can be translated into sample magnetic moment from knowledge of the relevant dimensions, or with somewhat greater accuracy, by cross-checking the low field Hall data with those obtained with a vibrating sample magnetometer (VSM).

Figure 3 shows the temperature-dependence of $H_{c2}$ as obtained from Hall probe micromagnetometry in the two principal directions. For $\mathbf{H}||\mathbf{c}$, there is good agreement with other recently-reported data,[9,10,11,12] and somewhat unexpectedly, $H_{c2}$ is rather lower than had been estimated from experiments on polycrystalline material. For fields in the *ab*-plane, the signal is an order of magnitude weaker, and the uncertainties correspondingly greater. We are limited to fields below 4 T, equivalent to temperatures above about 25 K, and in that range $H_{c2}^{ab}(T)$ appears to vary linearly with temperature. There are some reports of significant curvature of $H_{c2}^{ab}(T)$,[11,12] and indeed (see discussion below) there is no reason to expect all crystals to behave identically, but it is not yet clear whether the apparent differences are significant. The anisotropy $\gamma_{Hc2}$ of $H_{c2}$ between the two principal directions is close to 2, and we cannot discern any significant temperature dependence (figure 4).

We note that although the highest field to which *direct* systematic measurements of $H_{c2}$ has been made is 8 T, or about 20 K for fields in the *ab*-plane, attempts have been made at lower temperatures to utilise the G-L functional form of *M(H)* to obtain an extrapolated value of $H_{c2}$. However, these expressions are valid only within certain limits, e.g. large κ; furthermore, $MgB_2$ is not a simple superconductor. Hence such estimates should be treated with caution.

### II.B.  Measurement of $H_{c1}$

In an ideal pinning-free sample, $H_{c1}$ would be identifiable immediately from the peak at the end of the linear (perfectly diamagnetic) segment of the reversible *M(H)* curve. In real samples, bulk pinning, surface barriers, etc., contribute irreversible components of *M* that can easily overwhelm the reversible. Obviously, the cleaner the sample the better, but also small crystals are advantageous, in that $M_{rev}$ is size-independent, while $M_{irrev}$ scales with sample dimension. The essence of the experimental problem is to infer the reversible *M(H)* curve from the measured hysteretic loop; simple averaging of the upward and downward legs, i.e. assuming that for a given field $M_{irrev}$ changes sign but not magnitude with sweep direction, can be seriously misleading. However, one useful constraint on the analysis is that the reversible curve has to be linear in the Meissner phase with the correct slope, and must everywhere lie within the hysteretic loop.

Our studies utilise a double-axis transverse field VSM, in which the sample can be rotated and aligned *in situ*; it measures the components of the sample magnetic moment ***m*** (the magnetisation *M* is the moment per unit volume) parallel and perpendicular to the applied field.  The sample is first warmed above  $T_c$  and then cooled in zero applied magnetic field. We determine the field $H_p$ at which flux first penetrates the bulk of the sample by one of two methods, depending on the temperature. At low temperatures (where flux pinning is significant at low fields, so that $M_{rev}$ and $M_{irrev}$ are comparable in magnitude) $H_p$ is best found[13] by measuring the onset of flux trapping after successively larger field cycles of amplitude $H_{max}$. In a plot of the remnant (i.e. zero applied field) magnetic moment $m_{rem}$ after each cycle as a function of $H_{max}$ (figure 5), $H_p$ is readily identifiable as a sudden onset of a finite $m_{rem}$. Note that despite the presence of pinning, $H_p$ is close to the field at which the $m_{max}(H_{max})$ curve first deviates from linearity (because the reversible moment drops very sharply immediately above $H_{c1}$).

At temperatures above 25K and for the **H** || *ab* orientation, $m_{rem}$ becomes too small for the onset of flux trapping to be measured reliably. However, the pinning is then very weak, so that the $m_{max}(H_{max})$ curve itself shows a sharp deviation from linearity when flux first penetrates, so giving $H_p$ directly.

For **H** || *ab*, $H_p$ can be identified with $H_{c1}$ provided that the effects of Bean-Livingston (surface) barriers[14] and geometric barriers[15] are insignificant. Geometric barriers are not expected when the field direction is in the plane of a platelet sample, and so are irrelevant for **H** || *ab*. The Bean-Livingston barrier is strongly suppressed by surface irregularities (of which our sample has many) and would not be expected to play a significant role for either field direction. Finally, the Bean model for bulk pinning suggests that once $H_p$ is exceeded, $m_{rem}$ should increase quadratically as the flux front advances into the sample, in accord with our measurements.

With **H** || *c*, i.e. normal to the platelet, substantial demagnetising effects come into play. For an ellipsoid of demagnetising factor $N$, the field at the sample edge is enhanced by a factor $(1-N)^{-1}$. Our (somewhat irregular) platelet crystal, with a width $w$ to thickness $t$ aspect ratio of order 10, might be expected to behave roughly like an ellipsoid with $N \approx 1-\pi t/2w$. However, rather than relying on geometrical approximations, we use the fact that for a superconducting ellipsoid the initial magnetic susceptibility is also enhanced by the factor $(1-N)^{-1}$. We proceed from the measured low field slopes of the $m(H)$ loops in the two

field directions and obtain an effective demagnetising factor $N_{eff}$ such that $(1-N_{eff})^{-1} = (dm/dH)_{H//c}/(dm/dH)_{H//ab}=7.6\pm0.1$. We then equate $H_{c1}^{c}$ to $(1-N_{eff})^{-1}H_{p}$.

In both directions, the temperature-dependence of $H_{c1}$ is close to linear over the entire range from $T_c$ down to 5K (Figure 6), and shows no indication of levelling off, as it would within a simple BCS model.

We are aware of only one other attempt to measure $H_{c1}$ in MgB$_2$ single crystals over the entire temperature range. Zehetmeyer et al.[12] have investigated single-crystal low field behaviour using an ultra-sensitive SQUID magnetometer; although the magnetisation loops appear similar to ours, they detect initial flux penetration at much smaller fields. The inferred reversible curve would then violate the condition that it must lie within the hysteretic loop. One possibility is that, because the applied field is locally enhanced at corners and protuberances, very small amounts of flux are being seen entering those regions, rather than the onset of penetration into the bulk sample at the macroscopic $H_{c1}$. $H_{c1}^{ab}$ in a single crystal has been measured over a very limited temperature window by Sologubenko et al.[16] using the thermal conductivity as a probe; at 28K they find $H_{c1}^{ab} = 25\pm2$ mT, close to our value of 31±2mT.

## III.  GINZBURG-LANDAU PARAMETERS FOR MgB$_2$

For fields applied in the *c*-direction, the superconductive response is determined by the magnetic penetration depth $\lambda_{ab}$ associated with Meissner currents flowing in the *ab*-plane, and the superconducting coherence length $\xi_{ab}$, which reflects variation of the order parameter within the *ab*-plane, e.g. the radius of a vortex core. In the limit where $\kappa^c =(\lambda_{ab}/\xi_{ab})>>1$, G-L theory gives

$$\mu_0 H_{c1}^c = \left[\Phi_0 /\left(2\pi\lambda_{ab}^2\right)\right]\ln(\lambda_{ab}/\xi_{ab} + 0.5) \qquad (2)$$

and

$$\mu_0 H_{c2}^c = \Phi_0 / \left(2\pi\xi_{ab}^2\right) \quad (3)$$

Extrapolations of our data to zero temperature give $\mu_0 H_{c1}^c(0) = 0.28\pm0.01$T (figure 6) and $\mu_0 H_{c2}^c(0) = 3\pm0.5$T (figure 3). Equations (2) and (3) then yield $\lambda_{ab}(0) = 22\pm2$nm and $\xi_{ab}(0) = 10\pm0.2$ nm, hence $\kappa^c = 2.1\pm0.3$, violating the condition $\kappa \gg 1$.

Brandt has proposed a method for calculating the reversible $M(H)$ curve for $\mathbf{H} \parallel c$ for any value of $\kappa > 1/\sqrt{2}$;[17] our measured ratio of $H_{c1}^c/H_{c2}^c$ of $0.08\pm0.005$ (over the whole temperature range) implies a value of $\kappa^c$ of $3.4\pm0.2$. Also, $\kappa$ can be obtained by comparing the linear slope of $M(H)$ as $H$ tends towards $H_{c2}$ (Fig. 2(b)) with Brandt's curves; this method gives $\kappa^c = 3.6\pm0.2$ at 30K, rising slowly to $4.3\pm0.2$ at 5K.

For $\mathbf{H}\perp c$, screening currents have to flow in both the *ab*-plane and the *c*- direction, so that in equations 2 and 3, $\lambda_{ab}^2$ is replaced by $\lambda_{ab}\lambda_c$; likewise, $\xi_{ab}^2$ is replaced by $\xi_{ab}\xi_c$. Our measurements of $H_{c1}$ and $H_{c2}$ for this field direction translate to $\lambda_c(0) = 100\pm10$ nm and $\xi_c(0) = 5\pm0.2$ nm.

All these lengths are long compared with the unit cell dimensions of $MgB_2$, so that unlike the HTS materials, there is no danger of line-like vortices becoming fragmented into two-dimensional "pancake" vortices that are all too easily mobile. Furthermore, disruption of the regular structure, as at a grain boundary, usually extends no more than two or three unit cells, still quite short compared with $\xi_{ab}$ and $\xi_c$, so that such defects are unlikely to suppress superconductivity drastically, as they are prone to do in HTS.

Finally, the fluctuation parameter $Gi$, equal to $(k_B T_c / \mu_0 H_c^2 \xi_{ab}^2 \xi_c)^2$, which is a measure of thermal energy against the condensation energy of a coherence volume, is only about $10^{-6}$ in $MgB_2$, much smaller than the $10^{-2}$ that can be found in HTS, and approaching the $10^{-8}$ that is typical for LTS. Consequently, over almost all of the field-temperature plane, fluctuations are unimportant in $MgB_2$.

Thus in several key respects, $MgB_2$ is a much more promising superconductor than the HTS materials in terms of its *potential* for vortex pinning, and so for applications.

## IV. IMPACT OF THE GAP STRUCTURE

### IV.A. $H_{c2}$

There is now a consensus that the energy gap of $MgB_2$ has a complex structure, with at least two distinct gaps, as seen by point contact spectroscopy of the surface (Figure 7),[18] and corroborated by, for example, bulk heat capacity measurements.[19] The gap structure has recently been calculated,[20] and overall is in good agreement with the observed spectra.

In the present context, these aspects are very important, because they affect how the key parameters of $MgB_2$ may be influenced by the additional alloying and consequent scattering that is likely to be essential for high critical current performance.

The quasi-2 dimensional (2D) sheets of the Fermi surface arise from boron layer $p_x$ and $p_y$ σ orbitals, and provide high conductivity in the *ab*-plane, but little (none if they were purely 2D) in the *c*-direction. These states couple strongly to boron phonon modes, and drive the high transition temperature; they are responsible for the large gap of ~7 meV. The 3 dimensional (3D) sheets of the Fermi surface arise from boron layer $p_z$ π orbitals, and provide a less-anisotropic normal state conduction channel; they contribute the smaller, ~2.7 meV, gap that has lower spectral weight. There are no states at the Fermi level derived primarily from Mg orbitals.

Naively, it would be expected that elastic scattering between the sheets would readily hybridise the gap structure to yield a single average gap, corresponding to a lower $T_c$. An explanation of the survival of two distinct gaps and of a high $T_c$ has come from a detailed study of the impact of the band structure on superconductivity in $MgB_2$.[21] Essentially, the 2D and 3D sheets of the Fermi surface are rather well isolated from each other, because the symmetry of the structure and the orthogonality of the σ and π orbitals weaken the scattering matrix elements between them. Consequently, for any impurity, the intersheet scattering rates are much smaller than the intrasheet rates. Further, the 2D and 3D intrasheet rates may themselves be very different, and will depend upon the impurity species, particularly as to whether it lies in the Mg layer, or in the strongly-bonded B layer, the former being more likely. Similar considerations apply to the electron-phonon scattering rates.

Consequently, the measured resistivity just above $T_c$, where it is dominated by impurity scattering, is a poor guide to the scattering rates that are relevant to super-conducting properties. However, the estimated long mean free path in single crystals of $MgB_2$ obtained from a crude nearly-free electron model is so much longer than the G-L coherence lengths that the inference that these samples are in the clean limit must survive.

Bulk and thin film samples are commonly characterised in terms of the residual resistivity ratio (RRR), because of the uncertainties in the geometrical factor for translating measured resistance into resistivity. This too may not be straightforward, because while Matthiessen's Rule (the additivity of electron-impurity and electron-phonon scattering rates) should be a good approximation for each of the 2D and 3D sheets considered separately, it will not apply to the summed conductivities of the different channels.

In summary therefore, it should be possible to increase $H_{c2}$ substantially in $MgB_2$ by the introduction of scattering, so as to take the material out of the "clean" limit. Simple characterisations such as the resistivity will *not* be a definitive guide to the optimal scatterer to use to decrease ξ and increase $H_{c2}$.

### IV.B. Anisotropy

In the normal state, current flow in the *c*-direction of $MgB_2$ is dominated by the 3D sheets of the Fermi surface, but within the *ab*-plane 2D and 3D sheets contribute more equally. In the superconducting state, the sheets are coupled together through the electron-phonon interaction (and it is this coupling that allows the small gap to track the temperature-dependence of the large gap); consequently, the anisotropy $\gamma_{Hc2}$ may well be temperature-dependent, as some groups have reported. Also, the impact of impurity scattering on $\xi_c$, and so on $H_{c2}^{ab}$, will arise primarily from the increased collision rate on the 3D sheets, whereas $\xi_{ab}$ and $H_{c2}^{c}$ will be influenced by the collision rates on both 2D and 3D sheets. Consequently, as $MgB_2$ is loaded with scattering defects, $\gamma_{Hc2}$ may well alter, and possibly also its temperature-dependence; it would not be too surprising if different nominally-pure single crystals exhibit somewhat different behaviour in their anisotropy.

Although in G-L theory the anisotropy $\gamma_{Hc1}$ should be identical to $\gamma_{Hc2}$, there is no reason for the equality to survive in more complicated circumstances. We find (figure 4) that $\gamma_{Hc1}$ in a single crystal is temperature-independent, and in the overlapping temperature range is essentially equal to $\gamma_{Hc2}$, but that appears to be accidental. The physics of $H_{c1}$ is very different from that of $H_{c2}$, and rests on the cost of vortex entry into the bulk. We are not aware of any attempt to analyse this problem in detail in the context of two-band super-conductivity.

### V. PINNING

Since the coherence lengths are long compared with unit cell dimensions, isolated vacancies, substituents, interstitials, etc., contribute little directly to pinning, although they do of course enhance the scattering. Rather, as with LTS, extended defects (second-phase precipitates, intergrowths, dislocation tangles, grain boundaries) are required; the strained region around these defects may be important too. Because of the disparity between the relevant length scales, no simple correlation between resistivity and pinning is to be expected.

It should be said that the source of residual pinning in nominally "clean" material is unidentified. The pinning in single crystals is very much weaker than in polycrystalline material (Figure 8), implying many fewer extended defects. The fact that the transport scattering rates, as measured by the resistivities at $T_c$ of crystals and polycrystalline material, are very similar, around 1 $\mu\Omega$cm,[10] is not in any way a contradiction.

The dramatic changes in pinning behaviour after proton irradiation of polycrystalline fragments (Figure 9) are unlikely to arise from point defects alone. Possibilities are defect clustering, or interaction with pre-existing extended defects. Proton irradiation is not a scaleable process, but it provides a benchmark for other approaches. A number of groups have attempted the introduction of nanoparticles into bulk $MgB_2$ (which could be done on an industrial scale) to act as pinning sites, and these appear to be having some success,[22] but have not yet attained as good a field-dependence as the proton-irradiated material.

## VI.     CONCLUSIONS AND PROSPECTS FOR APPLICATIONS

For $MgB_2$ to be useful for high-current conductors, $J_C(B,T)$ must have its field dependence improved. The prospects are good for doubling the $H_{c2}$s from those found in single crystals, through the introduction of additional (targeted) scattering. Even heavier scattering may be counter-productive, as it is likely to induce mixing between the 2D and 3D sheets of the Fermi surface, and so reduce $T_c$; the anisotropy $\gamma_{Hc2}$ may be affected too. Empirically, effective pinning centres have been introduced into $MgB_2$ by a variety of routes, but they are far from being well-characterised. With better understanding, it should be possible to engineer defects of an optimised scale and density, and ones that can be fabricated by straightforward processing.

There appears to be no need (unlike HTS) to texture a $MgB_2$ conductor, and indeed in untextured material, a modest level of anisotropy may actually be helpful: at fields above $H_{c2}^c$, percolative paths will survive *via* grains that are oriented with their *c*-axes at sufficient angle to the applied field to remain superconducting, and so sustain $J_C$ up to high fields.

Little attention has been given to $H_{c1}$, but if the high values we have seen in single crystals can be attained in high quality textured films, the latter could be very attractive for microwave devices.

## VII.  ACKNOWLEDGMENTS

It is a pleasure to thank the organisers of the BOROMAG Workshop for their timely, successful and enjoyable initiative, and for the invitation to present this paper. Discussions with other participants at the Workshop have contributed significantly to the views presented here.

This research has been supported by the UK Engineering & Physical Sciences Research Council.

Figure 1  Critical current density $J_C$ of a MgB$_2$ sub-millimetre fragment of commercial (Alfa-Aesar 98% purity) powder as a function of magnetic field.[1]

Figure 2.  Sections of single crystal $m(H)$ loops in the vicinity of $H_{c2}$ at 33 K for **H**||*c* (a) VSM data, with a large sloping background (b) Hall micromagnetometry data; note that the signal is completely reversible with respect to field sweep direction.[7]

Figure 3  Upper critical fields of a MgB$_2$ single crystal as obtained directly from Hall probe micromagnetometry; measurements on other crystals from the same batch yield results that are identical within the error bars.[7]

Figure 4  Temperature dependences of the lower ($\gamma_{Hc1}$) and upper ($\gamma_{Hc2}$) critical field anisotropies of a MgB$_2$ single crystal.[7]

Figure 5  Low field response of a MgB$_2$ single crystal with $H \parallel ab$ at 10K. The sample has first been cooled in zero field, and then exposed to a series of field cycles of gradually increasing amplitude $H_{max}$. For each such cycle, both the moment $m_{max}$ at $H_{max}$ and the remanent moment (i.e. the moment in zero applied field) $m_{rem}$ are measured. Finally, a full hysteretic $m(H)$ loop is plotted out. The penetration field $H_p$ is well-defined, and for $H>H_p$, $m_{rem}$ increases in the manner expected from the Bean model.[7]

Figure 6  Lower critical fields of a MgB$_2$ single crystal obtained from magnetisation loops as in Figure 5.[7]

Figure 7  Point contact spectrum of a polycrystalline MgB$_2$ thin film at 4.2 K (data points). The curve is a fit to standard theory for two gaps, 2.4 meV, and 6.2 meV.[18]

Figure 8  Comparison at 10 K of magnetically-determined $J_C$ in a single crystal and a fragment of commercial MgB$_2$ powder.

Figure 9  Approaches to improving the quasi-exponential field-dependence of $J_C$ at 20 K through enhanced pinning. Proton irradiation at 1% displacements-per-atom (dpa)

reduces the slope by a factor of 3 from that in virgin material;[4] addition of yttria nanoparticles has a smaller, but still significant effect.[22]

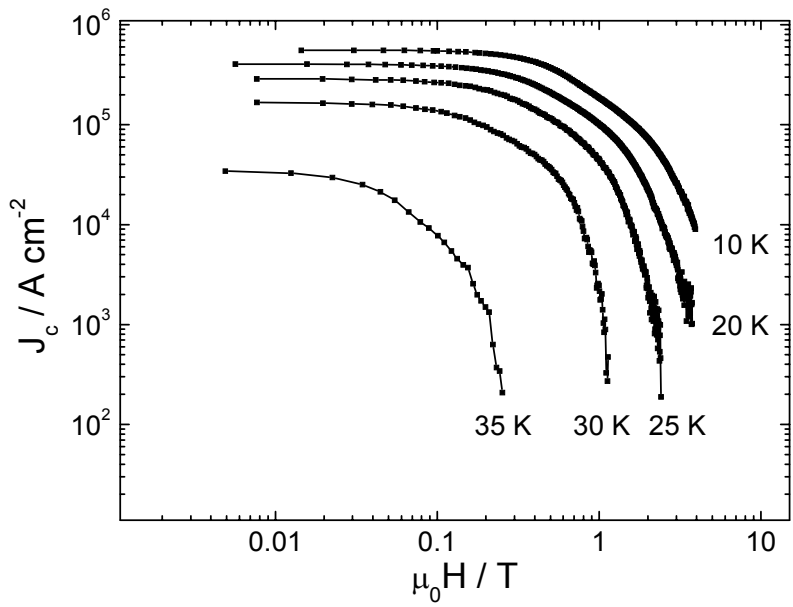

Figure 1    Critical current density $J_c$ of a MgB$_2$ sub-millimetre fragment of commercial (Alfa-Aesar 98% purity) powder as a function of magnetic field.

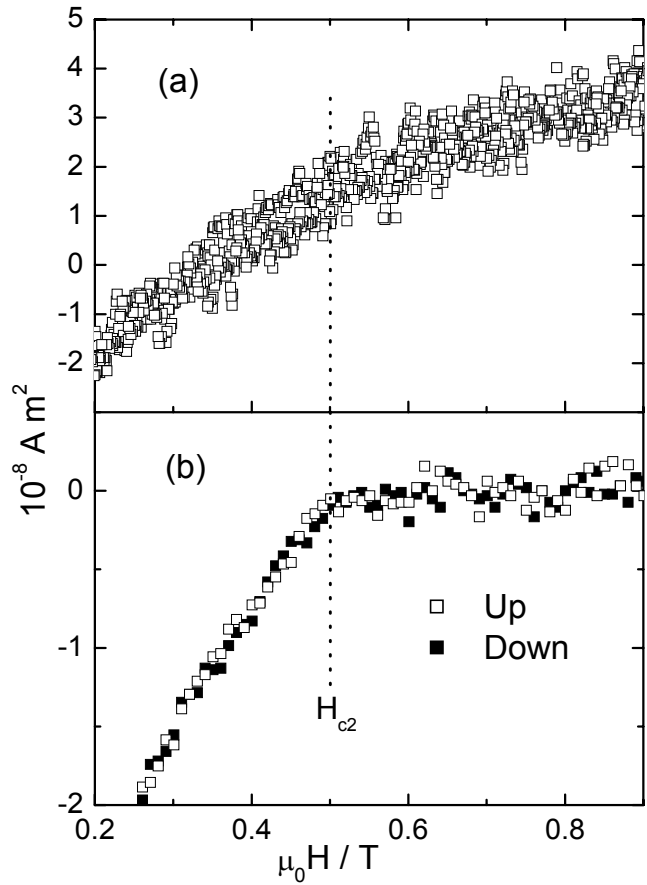

Figure 2  Sections of *m(H)* loops in the vicinity of $H_{c2}$ at 33 K for ***H***||*c* (a) VSM data, with a large sloping background (b) Hall micromagnetometry data; note that the signal is completely reversible with respect to field sweep direction.

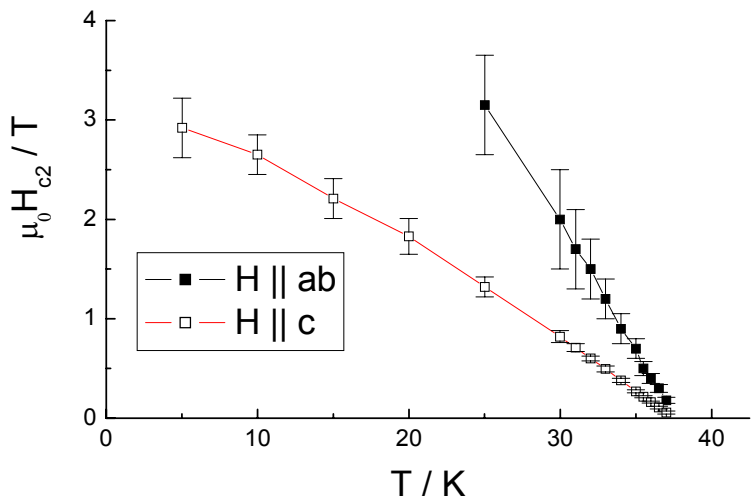

Figure 3  Upper critical fields of a $MgB_2$ single crystal as obtained directly from Hall probe micromagnetometry; measurements on other crystals from the same batch yield results that are identical within the error bars.

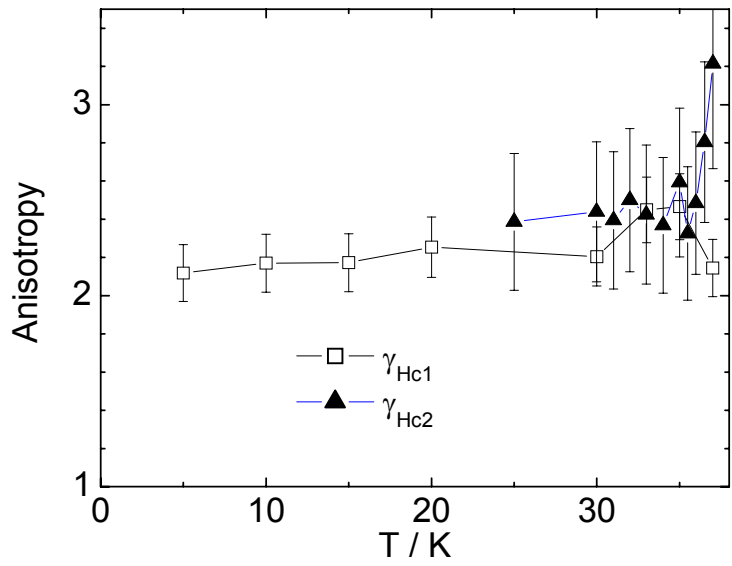

Figure 4    Temperature dependences of the lower ($\gamma_{Hc1}$) and upper ($\gamma_{Hc2}$) critical field anisotropies of a MgB$_2$ single crystal.

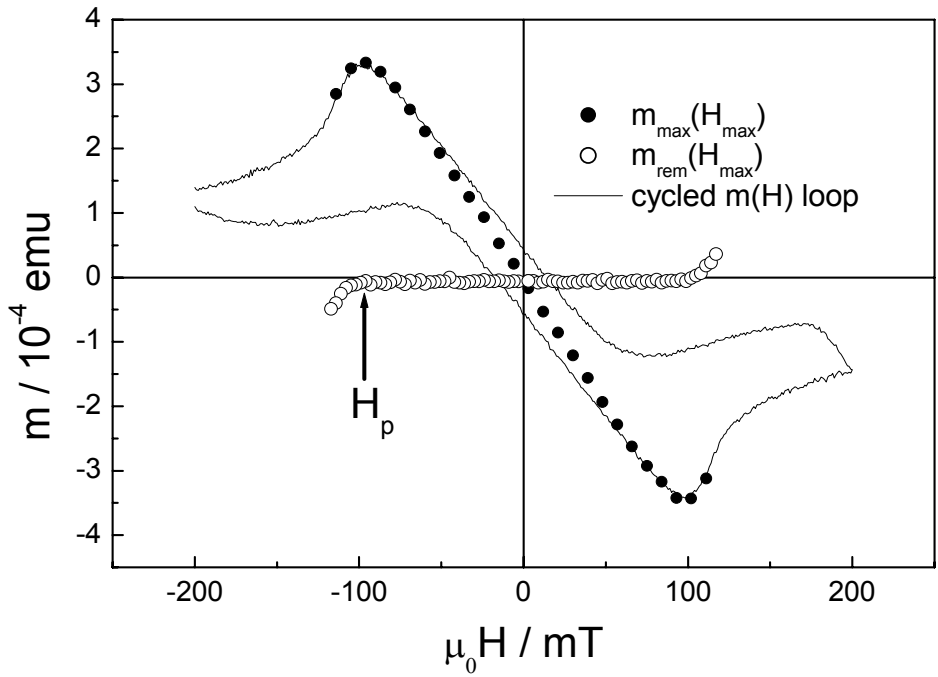

Figure 5  Low field response of a $MgB_2$ single crystal with $H \parallel ab$ at 10K. The sample has first been cooled in zero field, and then exposed to a series of field cycles of gradually increasing amplitude $H_{max}$. For each such cycle, both the moment $m_{max}$ at $H_{max}$ and the remanent moment (i.e. the moment in zero applied field) $m_{rem}$ are measured. Finally, a full hysteretic $m(H)$ loop is plotted out. The penetration field $H_p$ is well-defined, and for $H>H_p$, $m_{rem}$ increases in the manner expected from the Bean model.

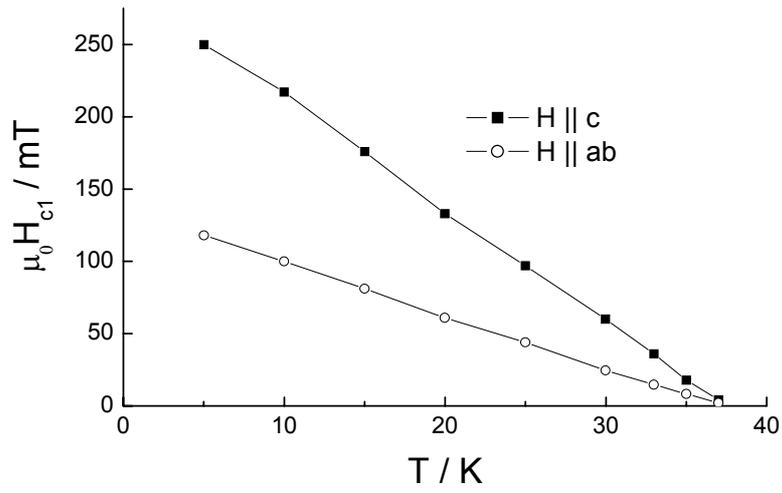

Figure 6    Lower critical fields of a $MgB_2$ single crystal obtained from magnetisation loops as in Figure 6.

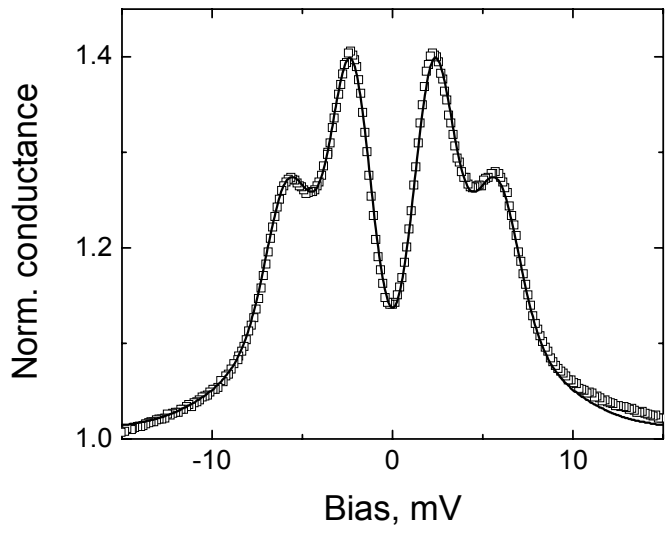

Figure 7    Point contact spectrum of a polycrystalline $MgB_2$ thin film at 4.2 K (data points). The curve is a fit to standard theory for two gaps, 2.4 meV, and 6.2 meV.

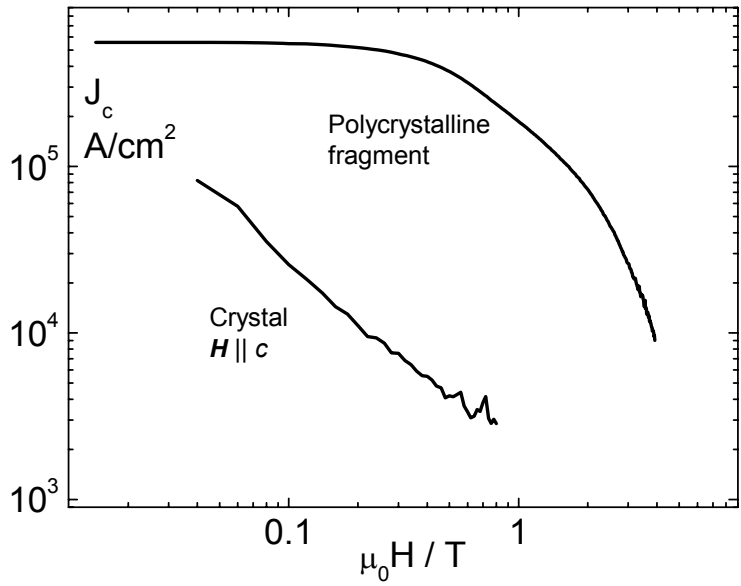

Figure 8   Comparison at 10 K of magnetically-determined $J_c$ in a single crystal and a fragment of commercial $MgB_2$ powder.

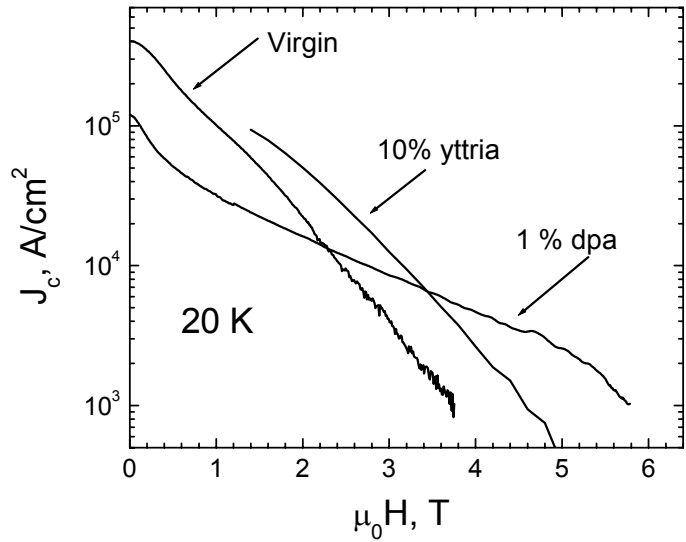

Figure 9  Approaches to improving the quasi-exponential field-dependence of $J_c$ at 20 K through enhanced pinning. Proton irradiation at 1% displacements-per-atom (dpa) reduces the slope by a factor of 3 from that in virgin material; addition of yttria nanoparticles has a smaller, but still significant effect.